\documentclass[prl,reprint,superscriptaddress,preprintnumbers]{revtex4-2}
\pdfoutput=1
\usepackage{amsfonts, amssymb, amsmath}
\usepackage{mathrsfs}
\usepackage[utf8]{inputenc}
\usepackage[russian,english]{babel}

\usepackage{color}
\definecolor{dark-gray}{gray}{0.20}
\definecolor{gray}{gray}{0.30}
\definecolor{light-gray}{gray}{0.80}
\definecolor{dark-red}{rgb}{0.7,0,0}
\definecolor{dark-green}{rgb}{0.1,0.4,0}
\definecolor{dark-blue}{rgb}{0.3,0.3,0.7}
\definecolor{light-blue}{rgb}{0.8,0.8,1}
\definecolor{blue}{rgb}{0,0,1}
\definecolor{red}{rgb}{1,0,0}
\definecolor{green}{rgb}{0,1,0}

\usepackage{hyperref}
\hypersetup{
	colorlinks=true,
	linkcolor=dark-blue,
	citecolor=dark-red,
	urlcolor=dark-blue,
	linktoc=page
}

\newcommand{\be}{\begin{equation}}
\newcommand{\ee}{\end{equation}}
\newcommand{\bea}{\begin{eqnarray}}
\newcommand{\eea}{\end{eqnarray}}

\begin{document}

\title{Black hole thermodynamics in natural variables:\\ the BTZ case}

\author{ Kiril Hristov}
\affiliation{Faculty of Physics, Sofia University, J. Bourchier Blvd. 5, 1164 Sofia, Bulgaria}
\affiliation{INRNE, Bulgarian Academy of Sciences, Tsarigradsko Chaussee 72, 1784 Sofia, Bulgaria}

\author{Riccardo Giordana Pozzi}
\affiliation{Dipartimento di Scienze Fisiche, Informatiche e Matematiche, Universit\`a di Modena e Reggio Emilia, via G. Campi 213/A, 41125 Modena, Italy}
\affiliation{INFN Sezione di Bologna, via Irnerio 46, 40126 Bologna, Italy}

$\;$
\begin{abstract}
It has been recently shown for a wide range of black hole solutions in 4 and 5 dimensions that it is useful to reorganize the conventional thermodynamics on the inner and outer horizons in terms of left- and right-moving variables. The insight of \cite{Hristov:2023sxg} regarding the split of variables was loosely inspired by the example of the BTZ black hole and the corresponding dual CFT$_2$ thermodynamics governed by the Cardy formula. Here we revisit the BTZ case and its generalizations in 3d gravity theories with higher derivative corrections, and show formally the applicability of the newly defined variables and their direct relation to the Cardy formula, which is not a priori guaranteed.   
\end{abstract}
%
%

%
\maketitle


\subsection{Introduction}
\label{sec:intro}
\vspace{-3mm}
The foundational principles of black hole thermodynamics were established during the 1970s through groundbreaking contributions by researchers such as Bekenstein, Bardeen, Hawking, and Gibbons \cite{Bekenstein:1973ur,Bardeen:1973gs,Hawking:1975vcx,Gibbons:1976ue}. Over subsequent decades, these principles have been extended and generalized in various ways. However, due to the absence of a comprehensive microscopic understanding provided by a quantum theory of gravity, these principles remain primarily an analogy to the laws of statistical thermodynamics. Such a quantum description is provided for asymptotically Anti-de Sitter (AdS) backgrounds embedded in string theory via the AdS/CFT correspondence \cite{Maldacena:1997re}. In the presence of supersymmetry many observables are protected and many microscopic calculations have successfully reproduces the macroscopic Bekenstein-Hawking entropy of the corresponding black holes. The task of characterizing realistic thermal black holes has instead proven to be significantly more challenging and exact microscopic calculations for such systems are still elusive. A notable successful exception from these hurdles is the case of AdS$_3$/CFT$_2$ correspondence because two-dimensional conformal symmetry is strong enough to fix the scaling of the microscopic density of states without any supersymmetry, resulting in the celebrated Cady formula, \cite{Cardy:1986ie}.\newline
\vspace{0cm} 
Considering the aforementioned insights, our current investigation aims to extend the program of \cite{Hristov:2023sxg}  and \cite{Hristov:2023cuo} to case of Ba\~nados-Teitelboim-Zanelli (BTZ) black holes \cite{Banados:1992wn,Banados:1992gq} asymptoting to AdS$_3$. \cite{Hristov:2023sxg} presented a fresh perspective on black hole thermodynamics by proposing a novel set of "natural" chemical potentials. These proposed potentials suggest a simpler microscopic explanation for thermal black holes. This approach is rooted in the intriguing observation that the laws governing black hole thermodynamics are not necessarily unique even within a single spacetime solution. Surprisingly, it is possible to define a set of chemical potentials and uphold a corresponding conservation law at each event horizon individually \cite{1979NCimB..51..262C}. For example, Hawking's original semi-classical calculation \cite{Hawking:1975vcx} applies equally to both the inner and outer horizons of a typical thermal black hole in an asymptotically flat 4-dimensional spacetime \cite{Cvetic:1997uw,Cvetic:1997xv,Wu:2004yk}. Similar features exist for additional event horizons in scenarios involving higher dimensions or a cosmological constant \cite{Cvetic:2018dqf}.\newline
\vspace{0cm}
Since the first law holds independently for multiple sets of chemical potentials, it becomes feasible to construct arbitrary linear combinations of the corresponding variables while preserving the same conservation law (known as the first law). This flexibility was previously utilized to define the so-called left- and right-moving entropies and temperatures \cite{Cvetic:1997uw,Cvetic:1997xv,Wu:2004yk}. More recently, \cite{Hristov:2023sxg,Hristov:2023cuo} expanded upon this concept by considering the corresponding left- and right-moving on-shell actions. Remarkably, these actions emerge as straightforward and fully explicit functions of the respective chemical potentials, a feature not exhibited by the free energies on the individual horizons.\newline
\vspace{0cm}
The construction in \cite{Hristov:2023sxg,Hristov:2023cuo} was also partially inspired by the natural split of left- and right-moving excitations of 2-dimensional conformal field theories (CFTs) compactified on a torus. We should however note that the black hole construction of natural variables made from combinations of potentials on multiple horizons is in no formal or clear way related to a CFT$_2$ symmetry, and the chosen terminology was based on an analogy rather than a clear match. The current investigation bridges this gap by elucidating the precise connection between the natural variables characterizing black holes, as applied to the BTZ case and its extensions featuring higher derivatives, see \cite{Detournay:2012ug}, and the corresponding left- and right-moving variables within CFT$_2$.

\subsection{The Cardy formula}
\label{sec:Cardy}
\vspace{-3mm}
Before moving to the investigation of the macroscopic black holes of interest, we briefly recall some basic facts about thermodynamics of 2-dimensional conformal field theories, see e.g.\ \cite{Kraus:2006nb,Hosseini:2020vgl} for reviews and further holographic context. We start by putting the field theory on a torus of modulus $\tau$,
\be
	\tau := \frac1{2 \pi}\, \left( \epsilon + i\, \beta \right)\ , 
\ee
which can be thought of a the complexified (inverse) temperature. Since the theory enjoys modular invariance, in the high-temperature limit $\tau \rightarrow 0$ it can be shown that the partition function (called elliptic genus) factorizes into independent holomorphic and anti-holomorphic contributions, which are in turn fully fixed. Considering the grand-canonical ensemble, we have
\be
	Z_{\text{CFT}_2} (\tau, \bar \tau) = Z_l (\tau)\, Z_r (\bar \tau)\ ,
\ee
with
\be
\label{eq:Cardy}
	\log Z_l (\tau) \approx \frac{i\, \pi}{12 \, \tau}\, c_l\ , \qquad \log Z_r (\bar \tau) \approx \frac{i\, \pi}{12 \, \bar \tau}\, c_r\ , 
\ee
where $\approx$ denotes the leading asymptotic behavior that we are interested in. Above, $c_l$ and $c_r$ are the central charges in the algebras of left-moving and right-moving excitations of the CFT$_2$, respectively. The above formula is the grand-canonical version of the Cardy formula, which ultimately fixes the asymptotic density of states. To achieve this, one needs to perform a Laplace transform of the partition function with respect to the number of left movers and right-movers $(n_{l,r} - c_{l,r}/24)$, conjugate to $\tau, \bar \tau$ respectively. At the leading order this amounts to a simple Legendre transform of the variables, leading to
\be
\label{eq:density}
	\log \rho (n_l, n_r) \approx 2 \pi\, \sqrt{\frac{c_l}{6}\, \left(n_l - \frac{c_l}{24} \right)} + 2 \pi\, \sqrt{\frac{c_r}{6}\, \left(n_r - \frac{c_r}{24} \right)}\ .
\ee
This is the more commonly used canonical version of the Cardy formula, that directly relates to the Bekenstein-Hawking entropy of BTZ black holes as we turn to discuss.

\subsection{BTZ in pure gravity}
\label{sec:GR}
\vspace{-3mm}
We consider the BTZ metric, \cite{Banados:1992wn,Banados:1992gq}, as a solution of the 3-dimensional Einstein-Hilbert action in presence of a cosmological constant, 
\be
\label{eq:BTZ-metric}
	{\rm d} s^2_{BTZ} = - U(r)^2 {\rm d} t^2 + U(r)^{-2} {\rm d} r^2 + r^2 ({\rm d} \phi +\frac{4 j}{r^2}\, {\rm d} t)^2\ ,
\ee
with $\phi$ is an angular coordinate and
\be
\label{eq:BTZ-metricfn}
	U(r)^2 = - 8 m + \frac{r^2}{l^2} + \frac{16 j^2}{r^2} = \frac{(r^2 - r_+^2) (r^2-r_-^2)}{r^2 l^2}\ ,
\ee
with $l$ setting the length scale of AdS$_3$, and $m, j$ the parameters governing the asymptotic mass $M$ and angular momentum $J$,
\be
\label{eq:MandJ}
	M = \frac{m}{G_N^{(3)}}\ , \qquad J = \frac{j}{G_N^{(3)}}\ .
\ee
The inner ($-$) and outer ($+$) black hole horizons are given by
\be
\label{eq:radii}
	r_\pm = \sqrt{2 l (l m + j)} \pm \sqrt{2 l (l m - j)} \ .
\ee
We can compute the Bekenstein-Hawking entropies and associated inverse temperatures on the two horizons, \cite{Detournay:2012ug}
\be
	S_\pm = \frac{\pi\, r_\pm}{2 G_N^{(3)}}\ , \quad \beta_\pm = \pm \frac{2 \pi l^2\, r_\pm}{r_+^2 - r_-^2}\ ,
\ee
as well as the angular velocities
\be
	\Omega_\pm = \frac{r_\mp^2}{l r_+ r_-}\ .
\ee
It is then easy to verify the first law of black hole thermodynamics holding separately at each horizon,
\be
	\beta_\pm\, \delta M = \delta S_\pm + \beta_\pm \Omega_\pm\, \delta J\ ,
\ee
such that the corresponding on-shell actions become
\be
\label{eq:IpmpureGR}
	I_\pm (\beta_\pm, \Omega_\pm) = \beta_\pm\,  M - S_\pm - \beta_\pm \Omega_\pm\, J = - \frac{\pi}{4\, G_N^{(3)}}\, r_\pm\ .
\ee

Let us just focus on the upper sign and rewrite the on-shell action as 
\be
\label{eq:IplusBTZ}
	I_+ = - \frac{3 l}{2 G_N^{(3)}} \left( \frac{2 \pi\, \sqrt{2 l (l m + j)} }{12\, l} + \frac{2 \pi\, \sqrt{2 l (l m - j)} }{12\, l} \right)\ ,
\ee
such that we obtain the equality
\be
\label{eq:match}
	I_+ = - \log Z_{\text{CFT}_2}\ ,
\ee
upon the identifications, see e.g.\ \cite{Kraus:2006wn},
\be
\label{eq:tauid}
	\tau = \frac{i\, l}{r_+ + r_-}\ , \qquad \bar \tau = \frac{i\, l}{r_+ - r_-}\ ,
\ee
and
\be
\label{eq:BrownH}
	c_l = c_r = \frac{3 l}{2 G_N^{(3)}}
\ee
Indeed the latter equality for the so-called Brown-Henneaux central charges needs to hold for the correct definition of AdS$_3$/CFT$_2$ as it can be derived from the asymptotic symmetries, \cite{Brown:1986nw}. The fact that the two central charges are equal is a defining characteristic of two derivative gravity with or without additional matter couplings. Similarly to \eqref{eq:match}, the entropy $S_+$ can be exactly equated to the logarithm of the density of states, \eqref{eq:density}.

\subsubsection{Natural variables}
Following \cite{Hristov:2023sxg}, we define
\bea
\label{eq:newvar}
\begin{split}
	\beta_{l,r} :=& \frac12\, (\beta_+ \pm \beta_-)\ , \qquad  \omega_{l,r} :=  \frac12\, (\beta_+ \Omega_+ \pm \beta_- \Omega_-)\ , \\  S_{l,r} :=&  \frac12\, (S_+ \pm S_-)\ , \qquad \qquad I_{l,r} :=  \frac12\, (I_+ \pm I_-)\ ,
\end{split}
\eea
leading to an alternative version of the first law,
\be
\label{eq:lrfirstlaw}
	\beta_{l,r}\, \delta M = \delta S_{l,r} + \omega_{l,r}\, \delta J\ ,
\ee
such that
\be
\label{eq:qsr-pureGR}
	I_{l,r} = \beta_{l,r} M - S_{l,r} - \omega_{l,r} J = - \frac{\pi \sqrt{2 l (l m \pm j)}}{4 G_N^{(3)}}\ .
\ee
It might now seem that we have doubled the number of chemical potentials that we are about to relate holographically, but in fact we find the following identities,
\be
\label{eq:betaomegaIdentity}
	\beta_l = -l\, \omega_l = \frac{\pi l^2}{2 \sqrt{2 l (l m + j)}}\ , \quad \beta_r = l\, \omega_r = \frac{\pi l^2}{2 \sqrt{2 l (l m - j)}}\ ,
\ee
such that we can rewrite
\be
	I_{l,r} = - S_{l,r} + \beta_{l,r} \frac{l M \pm J}{l}\ .
\ee
This means that in practice we again have only two independent variables, $\beta_l$ and $\beta_r$ in the left- and right-moving sectors, respectively. It is then easy to write down the on-shell actions in terms of their respective variables,
\be
	I_{l,r} = - \frac{\pi^2 l^2}{8 G_N^{(3)}\, \beta_{l,r}}\ , 
\ee 
obeying the identities
\be
\label{eq:lrconjugation}
	\frac{\partial I_{l,r}}{\partial \beta_{l,r}} = \frac{l M \pm J}{l}
\ee
It is then clear that under the identifications \eqref{eq:tauid}-\eqref{eq:BrownH},
\be
\label{eq:lrtau}
    \tau = \frac{i}{ \pi l}\, \beta_l\ , \quad  \bar \tau = \frac{i}{\pi l}\, \beta_r\ , 
\ee
leading to
\be
\label{eq:lrmatch}
	I_l (\beta_l) = - \log Z_l (\tau)\ , \quad I_r (\beta_r) = - \log Z_r (\bar \tau)\ ,
\ee
in agreement with \eqref{eq:match}. This illustrates the agreement between the left- and right moving sectors in gravity and their counterparts in field theory.

\subsubsection{Extremality}
The extremal limit for the BTZ black holes coincides with the supersymmetric (BPS) limit (when the theory is extended to supergravity),
\be
    M = \pm \frac{J}{l}\ ,
\ee
relating the two parameters describing the solutions, e.g.\ $j = \pm l\, m$. Both signs are equally meaningful, as they correspond either to $\beta_r \rightarrow \infty$ or $\beta_l \rightarrow \infty$, respectively. Choosing for clarity the upper sign above, we then find
\be
\label{eq:BPSlimit}
    I_l^\text{BPS} =  - \frac{\pi^2 l^2}{8 G_N^{(3)}\, \beta_{l}}\ , \qquad I_r^\text{BPS} = 0\ .
\ee
This is in agreement with the expectation that supersymmetry always imposes the vanishing of one of the sectors, observed generically in various examples in \cite{Hristov:2023sxg,Hristov:2023cuo}.

\subsection{BTZ in topologically massive gravity}
\label{sec:TMG}
\vspace{-3mm}
In topologically massive gravity (TMG), \cite{Deser:1981wh,Deser:1982vy}, the Einstein-Hilbert action in presence of a cosmological constant is suitably modified to include a gravitational Chern-Simons contribution parametrized by an additional coupling constant $1/\mu$ following the conventions of  \cite{Detournay:2012ug}. The addition of the topological term allows for the same black hole metric as above, \eqref{eq:BTZ-metric}-\eqref{eq:BTZ-metricfn}. The fact that we have a higher derivative correction however changes the resulting extrinsic potentials such as the asymptotic charges, entropy (calculated now via the Wald prescription \cite{Wald:1993nt}) and on-shell action. Taking this into account, see again \cite{Detournay:2012ug} for more details, the entropies for the outer and inner horizons are
\be
	S_{T, \pm} = \frac{\pi\, r_\pm}{2 G_N^{(3)}} + \frac{\pi\, r_\mp}{2 G_N^{(3)}\,\mu l}\
\ee
where $1/\mu$ is the Chern-Simons coupling defined as in \cite{Detournay:2012ug}. The conserved charges can be written as
\be
\label{eq:MandJforTMG}
    M_{T} = M+ \frac{J}{\mu l^2}\ , \, \quad J_{T} = J + \frac{M}{\mu}\ ,
\ee
where $M$ and $J$ are the asymptotic charges as defined in \eqref{eq:MandJ} and they become equivalent in the limit in which we go back to the pure gravity case, i.e. $\mu \rightarrow \infty$.

Again, the first law of black hole thermodynamics is verified for the two horizons 
\be
	\beta_\pm\, \delta M_{T} = \delta S_{T, \pm} + \beta_\pm \Omega_\pm\, \delta J_{T}\ ,
\ee
where $\beta_{\pm}$ and $\Omega_\pm$ are the inverse temperatures and the angular velocities as defined in the previous section. The on-shell actions,
\be
    \label{eq:I-TMG}
	I_{T, \pm} (\beta_\pm, \Omega_\pm) = \beta_\pm\,  M_T - S_{T,\pm} - \beta_\pm \Omega_\pm\, J_T\ ,
\ee
can be related to the ones in pure gravity, \eqref{eq:IpmpureGR}, by
\be
    I_{T, \pm} = I_\pm + \frac1{\mu l}\, I_{\mp}\ .
\ee
We again find an exact match between gravity and field theory, \eqref{eq:match}-\eqref{eq:tauid}, this time with the corrected central charges,~\footnote{Notice that here we have exchanged the definitions of left and right with respect to \cite{Detournay:2012ug}. This is a purely conventional choice for the purpose of notational consistency with the definition of the corresponding gravitational variables in \cite{Hristov:2023cuo,Hristov:2023sxg}.}
\be
\label{eq:BH-TMG}
    c_{l,r}  = \frac{3 l}{2 G_N^{(3)}}\Bigl(1 \pm \frac{1}{\mu l}\Bigr)\, ,
\ee
again in agreement with the Brown-Henneaux asymptotic analysis.

\subsubsection{Natural variables}
Given that the temperature and angular velocities do not change, the natural left- and right-moving variables in this case remain the same. Furthermore, following \eqref{eq:newvar}, we find that the new left- and right-moving on-shell actions satisfy the first law and relate to the entropies via the quantum statistical relation
\be
\label{eq:qsr-TMG}
    I_{T,\, l,r} = \beta_{l,r} M_T - S_{T,\, l,r} - \omega_{l,r} J_T\ ,
\ee
in analogy with \eqref{eq:qsr-pureGR}. It is also easy to verify that
\be
    I_{T, \, l,r} = \left(1 \pm \frac1{\mu l} \right)\, I_{l,r}\ ,
\ee
in terms of the analogous quantities in pure gravity. We finally arrive at
\be
    I_{T, \, l,r} = - \left(1 \pm \frac1{\mu l} \right)\, \frac{\pi^2 l^2}{8 G_N^{(3)}\, \beta_{l,r}}\ ,
\ee
leading to the identifications \eqref{eq:lrtau}-\eqref{eq:lrmatch} using the new central charges, \eqref{eq:BH-TMG}. Remarkably, we again obtain the analog of the conjugation relation \eqref{eq:lrconjugation}, 
\be
\label{eq:lrconjugation-tmg}
	\frac{\partial I_{T, \, l,r}}{\partial \beta_{l,r}} = \frac{l M_T \pm J_T}{l}\ ,
\ee
now with the corrected values of the asymptotic charges.

\subsubsection{Extremality}
In the TMG case the extremal and BPS limit corresponds to 
\be
    M_T = \pm \frac{J_T}{l}\ ,
\ee
which at the level of the basic black hole parameters remains the same, $j = \pm l\, m$. This again means a vanishing temperature, $\beta_{l,r} \rightarrow \infty$. Choosing the upper sign we then find
\be
    I_{T,\, l}^\text{BPS} =  - \left(1 + \frac1{\mu l} \right)\, \frac{\pi^2 l^2}{8 G_N^{(3)}\, \beta_{l}}\ , \qquad I_{T, \, r}^\text{BPS} = 0\ ,
\ee
in analogy to \eqref{eq:BPSlimit}.

\subsection*{Generalizations and conclusions}
Through the examples presented above, we have demonstrated the correspondence between the microscopic variables in field theory and the gravitational formulation of left- and right-moving sectors proposed independently in \cite{Hristov:2023sxg}. This equivalence is non-trivial, as the gravitational construction initially involves doubling the chemical potentials to account for thermodynamics on both the inner and outer horizons. However, in the BTZ case, this doubling is counteracted by the additional relations \eqref{eq:betaomegaIdentity}, resulting in each sector featuring only a single independent chemical potential, associated with the modular parameter in the dual theory.

Remarkably, these observations can be formally extended to other theories incorporating higher derivative corrections, such as new massive gravity \cite{Clement:2009gq} and Lagrangians of the form $f(g_{\mu\nu}, R_{\mu\nu})$ as demonstrated in \cite{Saida:1999ec}. In these theories, only an overall normalization factor is found, maintaining the equality of the Brown-Henneaux central charges. It is noteworthy that the TMG case discussed earlier falls outside this category due to the gravitational Chern-Simons term, which disrupts the symmetry between the two sectors. The fact that our gravitational variables consistently yield accurate results suggests their broad applicability across these theories.

Additionally, it is essential to recognize another class of AdS$_3$ vacua and their BTZ-like quotients in higher derivative theories known as "warped" AdS \cite{Anninos:2008fx,Chow:2009km}. In this scenario, the dual field theory deviates from a standard two-dimensional CFT and instead exhibits only a single copy of the Virasoro algebra, \cite{Detournay:2012pc,Compere:2013bya}. Consequently, the corresponding partition function and Cardy formula analog only feature a single sector (the right-moving one in our conventions), implying the absence of the left-moving sector. It is intriguing to note that our gravitational construction precisely aligns with the dual variables once again without any changes in the gravitational construction. We plan to report on this observation elsewhere.

\subsubsection*{Acknowledgements}
\vspace{-3mm}
The study of K.H. is financed by the European Union- NextGenerationEU, through the National Recovery and Resilience Plan of the Republic of Bulgaria, project No BG-RRP-2.004-0008-C01.

\bibliographystyle{apsrev4-2}
\bibliography{btz.bib}

\end{document}